\documentclass[aps,prd,nofootinbib,onecolumn,groupedaddress,showpacs,showkeys]{revtex4}

\usepackage{graphicx}
\usepackage{epsfig,latexsym,amssymb}
\usepackage{amsmath}
\usepackage{empheq}
\usepackage{bm}
\usepackage{slashed}
\usepackage{subfigure}
\usepackage{comment}
\usepackage[dvipsnames]{xcolor}

\newcommand{\be}{\begin{equation}}
\newcommand{\ee}{\end{equation}}
\newcommand{\bea}{\begin{eqnarray}}
\newcommand{\eea}{\end{eqnarray}}
\newcommand{\nn}{\nonumber}

\def\eq#1{{Eq.~(\ref{#1})}}
\def\fig#1{{Fig.~\ref{#1}}}
\newcommand{\ben}{\begin{eqnarray*}}
\newcommand{\een}{\end{eqnarray*}}

\usepackage[normalem]{ulem}  
\renewcommand\sout{\bgroup \color{red} \ULdepth=-.5ex \ULset}

\renewcommand{\mathtt}[1]{{#1}}

\begin{document}

\title{Non-perturbative renormalization of the average color charge and multi-point correlators 
of color charge from a non-Gaussian small-$x$ action}
\author{Andre V. Giannini$^{1,2}$, and Yasushi Nara$^1$}
\affiliation{ 
$^1$ Akita International University, Yuwa, Akita-city 010-1292, Japan \\
$^2$ Instituto de F\'{i}sica, Universidade de S\~ao Paulo,
Rua do Mat\~ao 1371,  05508-090 S\~ao Paulo-SP, Brazil \\
}

\begin{abstract}
The McLerran-Venugopalan (MV) model is a Gaussian effective theory
of color charge fluctuations at small-$x$ in the limit of large valence 
charge density, {\it i}.{\it e}., a large nucleus made of uncorrelated 
color charges. 
In this work, we explore the effects of the first non-trivial 
(even C-parity) non-Gaussian correction on the color charge 
density to the MV model (``quartic" term) in SU(2) and SU(3) 
color group in the non-perturbative regime. 
We compare our (numerical) non-perturbative results to (analytical) 
perturbative ones in the limit of small or large non-Gaussian fluctuations.
The couplings in the non-Gaussian action, $\bar\mu$ for the 
quadratic and $\kappa_4$ for the quartic term, need to be 
renormalized in order to match the two-point function in 
the Gaussian theory. 
We investigate three different choices for the renormalization  
of these couplings: 
i) $\kappa_{4}$ is proportional to a power of $\bar\mu$; 
ii) $\kappa_4$ is kept constant and 
iii) $\bar\mu$ is kept constant.
We find that the first two choices lead to a scenario where 
the small-$x$ action evolves towards a theory dominated by 
large non-Gaussian fluctuations, regardless of the system size, 
while the last one allows for controlling the deviations from 
the MV model.
\end{abstract}

\maketitle

\section{Introduction}

As dynamic emission of soft gluons (over-)populates the phase space at high
energies, hadrons may be described as a classical system.
Such description is provided by the Color Glass 
Condensate (CGC) effective field theory~\cite{CGC.review.new,CGC.effective.theory}, where calculations 
rely on a scale separation: large-$x$ (``valence") partons act as a randomly distributed static color sources 
$\rho$ that generate the dynamical, short-lived, small-$x$ gluons. 
Due to the stochastic nature of the color charges, the resulting small-$x$
field, obtained by solving Classical Yang-Mills equations for a particular source configuration, 
must be averaged over a given ensemble $W_{Y}[\rho]$ of color charges. 
Therefore, any quantity of interest is obtained as the following expectation value,
\be\label{eq:coloravg}
\langle \mathcal{O}[\rho] \rangle_{Y} = \frac{\int [d\rho]\, W_{Y}[\rho]\, \mathcal{O}[\rho]}{\int [d\rho]W_{Y}[\rho]}\,,
\ee
where $Y=\log(x_0/x)$, with $x_{0}\sim 0.01$, denotes the rapidity variable.

Quantum corrections for $\langle \mathcal{O}[\rho] \rangle_{Y}$ due to the evolution in 
rapidity/energy are taken into account via the Wilsonian renormalization group equation 
for $W_{Y}[\rho]$ known as JIMWLK equation~\cite{JalilianMarian:1996xn,JalilianMarian:1997jx,JalilianMarian:1997gr,
JalilianMarian:1997dw,JalilianMarian:1998cb,Kovner:1999bj,Kovner:2000pt,Iancu:2000hn,Iancu:2001ad,Ferreiro:2001qy}.
Solving such an evolution equation is an initial value problem;
it requires an initial distribution of color charges as input. 
For an infinitely large nucleus made of uncorrelated color charges,
it is possible to show that $W_{0}[\rho]$ is a Gaussian,
which is known as the McLerran-Venugopalan (MV) model~\cite{CGC.Raju.McLerran},
and it is widely employed in CGC calculations.

In reality, however, the number of color charges is finite 
and their distribution should deviate from a Normal one. 
Such deviation should occur even in the absence of quantum 
corrections and also for large nuclei~\cite{Lam:2001ax}, 
as the finiteness of color charges by itself introduce correlations.
Therefore, the initial condition for the evolution equation is not 
necessarily a Gaussian.
It is known that a Gaussian distribution is not a solution of the JIMWLK evolution 
equation~\cite{JalilianMarian:1997gr}, and the small-$x$ 
evolution generates non-quadratic terms (in the color charge $\rho$) 
even if one starts with a Gaussian distribution of color charges.
Non-Gaussian contributions were indirectly studied within 
JIMWLK evolution. Starting with a Gaussian initial condition 
(MV model), it was found that the small-$x$ evolution appears to 
preserve the Gaussianity of the initial color charge distribution 
for two specific configurations (``line" and ``square" 
configurations) of the correlator of four Wilson lines 
in~\cite{Dumitru:2011vk}. At the same time, the product 
of the correlator of two and four Wilson lines, which 
is present in the cross-section for di-hadron production in 
proton-nucleus collisions~\cite{Marquet:2007vb}, 
has shown deviations from analytical expressions obtained in 
the Gaussian approximation~\cite{Dominguez:2011wm} in the 
saturation region. 
It is still unknown what happens if one starts the 
evolution with a non-Gaussian initial condition instead of 
considering the MV model.
It was pointed out in~\cite{Lappi:2015vta} that
the small-$x$ evolution may introduce 
non-Gaussian contributions in some observables such as the 
azimuthal anisotropies, $v_n$.

Corrections to the MV model for SU(3) color group have already been calculated 
in the literature~\cite{Jeon:2004rk,Dumitru:2011zz} up to the fourth-order 
in the color charges. The resulting non-Gaussian weight function has then 
been used to perform perturbative calculations in the dilute 
regime~\cite{Dumitru:2011zz,Dumitru:2011ax,Dumitru:2012tw}, where 
the corrections to the MV model are assumed to be small.
The impact of a non-Gaussian weight function on observables 
has not been investigated in details yet, but it is expected that it 
could lead to a better representation of the initial conditions for 
proton-proton and proton-nucleus collisions. 
Moreover, such higher-order terms may contribute to 
experimental observables in different physical processes, 
such as multi-particle correlations in nuclear 
collisions~\cite{Dumitru:2011zz,Kovner:2010xk}, 
di-jets produced in proton-proton and 
proton-nucleus~\cite{Marquet:2007vb,Dominguez:2011wm}
and inclusive Deep Inelastic Scattering structure functions, $F_L$ and $F_2$, which 
can be related to the forward scattering amplitude of a 
quark-antiquark pair~\cite{GolecBiernat:1998js}.

In this work, we present a first study of the effects of non-Gaussian 
corrections to the MV model on multi-point correlations of color charges
in the fully non-perturbative regime.
Specifically, we investigate 
three different renormalization schemes for determining the 
couplings of the non-Gaussian small-$x$ action. 
Our calculations will be done in lattice regularization and 
carried out for $Y=0$; therefore, they may be 
used as initial conditions for the renormalization group equations
to go beyond the Gaussian approximation in the CGC effective theory.
We shall show below that one of these renormalization schemes allows us
to control the deviations from the MV model as one approaches the continuum 
while the other two lead to a small-$x$ action that evolves towards 
a theory dominated by strong non-Gaussian fluctuations
regardless of the system size.

In the next section,
we briefly present the Gaussian and non-Gaussian effective weight functions
which are used to take an average over the color sources
in the CGC approach. 
We then present perturbative results in the limit of small as well as 
large non-Gaussian fluctuations,
and compare them to non-perturbative calculations,
which includes all orders of $1/\kappa_4$
(see \eq{eq:quarticAction} for the definition of $\kappa_4$).

\section{Weight functions for color charge average}

The central limit theorem applies
in the high density limit for color charge density and 
the absence of correlations between color charges at 
different coordinates~\cite{Lam:2001ax}.
Then,
$W[\rho]$ is given by the MV model~\cite{CGC.Raju.McLerran}
\be
W_{MV}[\rho_{x_\perp}] = \exp \bigg\{ - \int \, d^2 {x_\perp} 
{\delta^{ab} \,  \rho^a_{x_\perp} \rho^b_{x_\perp} \over 2 \mu^2} \bigg\}~,  \label{eq:MVAction}
\ee
where $\mu^{2}$ represents the average color charge squared per unit area per
color degree of freedom, and 
$\rho^{i}_{{x_\perp}}\equiv\rho^{i}({x_\perp})$ is the color charge
density at a 
given transverse coordinate $x_\perp$. 
In this case, the two-point function of color charge density is the 
only non-trivial correlator: any higher-order $n$-point function 
($n = 4,6,8...$) can be factorized into a product of $n/2$ two-point 
functions.

We shall consider deviations from a Gaussian weight due to finite number of
color sources.
Non-Gaussian corrections to \eq{eq:MVAction} for SU($N_c$) color group, where $N_c \leqslant 3$,  have been calculated in the 
literature~\cite{Jeon:2004rk,Dumitru:2011zz} up to the forth-order in the color 
charges:
\bea
W[\rho_{x_\perp} ] &\simeq&  \exp\bigg\{ - \int \, d^2 x_\perp \bigg[
{\delta^{ab} \,  \rho^a_{x_\perp} \rho^b_{x_\perp} \over 2 {\bar\mu}^2} - {d^{abc}\, \rho^a_{x_\perp} \rho^b_{x_\perp}
	\rho^c_{x_\perp} \over \kappa_3 } +
{\delta^{ab} \delta^{cd} +  \delta^{ac} \delta^{bd} + \delta^{ad}
	\delta^{bc}  \over \kappa_4}
\rho^a_{x_\perp} \rho^b_{x_\perp} \rho^c_{x_\perp} \rho^d_{x_\perp} \bigg] \bigg\} \nn \\  \label{eq:quarticAction}
&=& 
\exp\bigg\{ - \int \, d^2 x_\perp \bigg[
{  \rho^a_{x_\perp} \rho^a_{x_\perp} \over 2 {\bar\mu}^2} - {d^{abc}\, \rho^a_{x_\perp} \rho^b_{x_\perp}
	\rho^c_{x_\perp} \over \kappa_3 } +
{3 \over \kappa_4}
\rho^a_{x_\perp} \rho^a_{x_\perp} \rho^b_{x_\perp} \rho^b_{x_\perp} \bigg] \bigg\}\,,
\eea
where $\delta^{ij}$ is the Kronecker's delta, $d^{abc}$ is the 
symmetric tensor in the SU(3) Lie algebra~\cite{Jeon:2004rk} 
and $\bar\mu^{2}$ is the average color charge squared, 
$\kappa_3$ and $\kappa_{4}$ represent the couplings from 
the first odd C-parity (``cubic" term) and even C-parity 
(``quartic" term) corrections to the MV model. 
When non-Gaussian corrections to the MV model are small, the couplings 
in \eq{eq:quarticAction} can be written as~\cite{Dumitru:2011zz}: 
\bea
\bar\mu^2 \equiv {g^2 A \over 2 \pi R^2} \,, \label{eq:NGcouplings}
\qquad\qquad 
\kappa_3 \equiv 3{g^3 A^2 \over (\pi R^2)^2} \,  = {12 \bar\mu^4 \over g} \,, 
\qquad\qquad 
\kappa_4 \equiv 18{g^4 A^3 \over (\pi R^2)^3} \,   = {144 \bar\mu^6 \over g^2}\,,
\label{eq:nongausscouplingSU3}
\eea
where $A$ represents the mass number and $R$ the radius of 
the system of interest. For SU(2), $\kappa_{4}$ is given by
\bea
\kappa_4 \equiv 6{g^4 A^3 \over (\pi R^2)^3} \, = {48 \bar\mu^6 \over g^2}.
\label{eq:nongausscouplingSU2}
\eea
The factor in \eq{eq:nongausscouplingSU2} can be verified in two different ways: by following 
the calculation in~\cite{Jeon:2004rk} but including higher-order 
contributions in the Taylor expansion of $G_{k:s}$ in
their Eq. (19) and by writing the quartic Casimir in 
Eq. (12) of~\cite{Dumitru:2011zz} for SU(2). 
We note that the small-$x$ action for SU(2) symmetry group 
does not have cubic (``odderon") term as the symbol $d^{abc}$  is zero.

In what follows, we only consider the quartic term in 
the SU(3) case, leaving the study of the cubic term in the future.
While corrections (at perturbative level) due to the 
inclusion of the cubic term are expected at order $1/\kappa_3^2$ for 
SU(3)~\cite{Dumitru:2011zz}, our results for SU(2) are exact up to all 
orders in $1/\kappa_4$.

The coupling in the quadratic and the quartic term 
in \eq{eq:quarticAction}, $\bar\mu$ and $\kappa_4$,
are not chosen freely. 
It is required that the inclusion of non-Gaussian corrections does 
not impact any quantity depending solely on the correlator of 
two-color charges, 
$\langle \rho^a_{x_{\perp}}\rho^b_{y_{\perp}}\rangle$, 
since this is a quantity determined by the quadratic part 
of the small-$x$ action. 
Thus, one needs to renormalize the couplings in the non-Gaussian 
action to match the two-point function of color charges 
from the Gaussian theory.
In this way, the two couplings are related to each other 
and one more condition is needed to uniquely fix them.
We consider three possible ways to fix these couplings.
One option is to keep $\kappa_4 / \bar\mu^6\equiv \lambda$ constant.
In principle, $\lambda$ can be fixed to any (positive) value. 
Motivated by the expression for $\bar\mu^2$ in \eq{eq:NGcouplings},
we take
$\lambda=\gamma / g^2$, where $\gamma= 48\, (144)$ for 
SU(2) (SU(3)).
A second option is to keep $\kappa_{4}$ constant.
The parametric dependence shown 
in \eq{eq:nongausscouplingSU3} and \eq{eq:nongausscouplingSU2} 
is no longer valid in this renormalization scheme.
The third option is to control the deviation from the MV model by fixing
the parameter $Z$ defined by
\be
Z = \frac{\mu^2}{\bar\mu^2}\,.
\label{eq:renormparam}
\ee
We shall show that the first two renormalization 
schemes lead to a theory dominated by non-Gaussian 
fluctuations independent of the system size.

\section{(Semi-)Analytical results for color average in the transverse lattice}

This section presents the expressions for the correlators of two- and
four-color charges for different approximations.
The first approximation is to assume that the quartic term in
\eq{eq:quarticAction} is small~\cite{Dumitru:2011zz}.
The second considers a limit of large non-Gaussian corrections,
in which the quartic term is large.
Otherwise stated, all expressions will be presented in lattice regularization by 
approximating the two-dimensional transverse space by $N_s^2$ lattice sites with 
lattice spacing $\mathtt{a}$.

For a weight function which only involves the product of square power of color charges, 
as in the case for SU(2) and SU(3) without the cubic term, 
one can calculate the color average in~\eq{eq:coloravg} on a lattice by evaluating
\be
\langle \mathcal{O} \rangle = \frac{\int(\prod_{x} \prod_a  d\rho^a_{x})\, \mathcal{O} \, e^{-\sum_y W_{y}} }{\int(\prod_{x}\, \prod_a d\rho^a_{x})\, e^{-\sum_y W_{y}}}
= \frac{\int(\prod_a  d\rho^a_{x})\, \mathcal{O}_{x} \, e^{-W_{x}} }{\int(\prod_a d\rho^a_{x})\, e^{-W_{x}}}\, 
= \frac{\int  dr\, r^{N_{c}^2-2} \, \mathcal{O}_r \, e^{-W_{r}} }{\int dr\, r^{N_{c}^2-2} \, e^{-W_{r}} }\,,
\label{eq:coloravglattice}
\ee
where $W_r$ is defined as
\be
W_r = \frac{a^2r^2}{2\bar\mu^2}+\frac{3a^2r^4}{\kappa_4}\,.
\label{eq:Wlattice}
\ee
The second equality in \eq{eq:coloravglattice} is 
obtained by assuming that $\mathcal{O}$ is a 
local operator. 
As discussed below, the correlators of two- and 
four-color charges will be affected by non-Gaussian 
corrections when calculated locally.
In such configuration, the functional 
integral becomes an integral over the color charges at a single site. 
The rightmost result is then obtained after using spherical coordinates
in $N_{c}^2 - 1$ dimensions, which factor out any angular dependencies.

The requirement that any quantity depending only on the two-point function of color 
charges remains unchanged introduces the following constraint\footnote{To avoid cluttered 
notation, from now on, $x$ denotes a point in the transverse lattice, not the fraction of 
momentum carried by produced gluons; moreover, we omit the $\perp$ notation in the transverse coordinates.}
\be
\langle \rho^a_{x}\rho^b_y\rangle_{\text{non-Gaussian}} = \langle \rho^a_{x}\rho^b_y\rangle_{\text{MV model}} = 
\frac{\delta^{ab}\delta_{xy}}{\mathtt{a}^2}\mu^2\,, 
\label{eq:renormalizationMu}
\ee
where $\mathtt{a}^2$ represents the area of a lattice cell, $x$ and $y$ are discrete points
in the transverse lattice, and $\delta_{xy}/{\mathtt{a}^2}$ is
the lattice counterpart of the Dirac's delta, $\delta(x-y)$; from here on,
we use the shorthand notation ``NG" to denote results obtained using the 
non-Gaussian weight function. 
Thus, one of the couplings in the non-Gaussian weight function is chosen in 
order to satisfy \eq{eq:renormalizationMu}.

As noted above, local operators do not present a spatial dependence over 
a two-dimensional lattice and the integral is only over the color space.
Then the correlator of two-color charges (${\mathcal O}_x = \rho_x^a\rho_y^b$) is given by
\be
\langle \rho^a_{x}\rho^b_{y} \rangle_{\text{NG}}
= \delta^{ab}\delta_{xy}\,\frac{\mu^2\sqrt{X}}{Z\,a^2} 
\frac{ U\left(\frac{1}{4} \left(N_c^2 +1\right),\frac{1}{2},X\right)}{U\left(\frac{1}{4} \left(N_c^2-1\right),\frac{1}{2},X\right)}
= \delta^{ab}\delta_{xy}\,\frac{\sqrt{\kappa_{4}}}{4\,\sqrt{3}\,a} 
\frac{ U\left(\frac{1}{4} \left(N_c^2 +1\right),\frac{1}{2},X\right)}{U\left(\frac{1}{4} \left(N_c^2-1\right),\frac{1}{2},X\right)}
\label{eq:rho2nonGauss}
\ee
where $X = a^2\, \kappa_4 / 48 \bar\mu^4$ and
\be
U(\alpha,\beta,\omega)= \frac{1}{\Gamma(\alpha)} \int_0^\infty e^{-\omega t}\, t^{\alpha-1}\, (1+t)^{\beta-\alpha-1}\,dt
\ee 
denotes the Tricomi's confluent hypergeometric function; $\Gamma(\alpha)$ 
is the Gamma function.
The condition \eq{eq:renormalizationMu} for SU($N_c$) is given by
\be
\frac{\bar\mu^2\sqrt{X}}{a^2}
\frac{U\left(\frac{1}{4} \left(N_c^2 +1\right),\frac{1}{2},X\right)}
       {U\left(\frac{1}{4} \left(N_c^2-1\right),\frac{1}{2},X\right)}
=
\frac{\mu^2}{\mathtt{a}^2}\,.
\label{eq:renormMuSU2}
\ee
The four-point function of color charges can also be 
expressed in terms of the Tricomi's confluent hypergeometric function:
\be
\langle \rho^a_{x}\rho^a_{x}\rho^c_{x}\rho^c_{x} \rangle_{\text{NG}} = 
\left(N_c^4-1\right)\,
\frac{\bar\mu^4}{a^4}\frac{XU\left(\frac{1}{4} \left(N_c^2+3\right),\frac{1}{2},X\right)}{U\left(\frac{1}{4} \left(N_c^2-1\right),\frac{1}{2},X\right)}\,.
\label{eq:rho4NonGauss}
\ee

We finish this section by summarizing the four-point function of 
color charges, $\langle\rho_x^a\rho_y^b\rho_u^c\rho_v^d\rangle$
in the MV model.
Calculating the color average in~\eq{eq:coloravglattice} 
with the Gaussian ensemble results in 
\be
\langle \rho^a_{x}\rho^b_y \rho^c_u\rho^d_v \rangle_{\rm MV} = \mu^4
\left[ \,
\delta^{ab}\delta^{cd}
\delta(x-y)\delta(u-v)
+ 
\delta^{ac}\delta^{bd}
\delta(x-u)\delta(y-v)
+
\delta^{ad}\delta^{bc}
\delta(x-v)\delta(y-u)\, 
\right]\,.
\label{eq:rho_4_MV_lattice}
\ee
We point out that the color factor multiplying $\mu^4$ is dependent on 
the configuration.
We have a factorizable configuration
in the lattice configuration that each pair of color charges sit at 
different sites ({\it i}.{\it e}. $x=y$, $u=v$ but $x\neq u$).
Each one of the two-point function contributes
with a factor of $N_c^2-1$ after contracting color indexes. 
The coefficient of the correlator of four-color charges then evaluates to 
\be
\langle\rho_x^a\rho_x^a\rho_u^c\rho_u^c\rangle_{\rm MV} = 
\langle\rho_x^a\rho_x^a\rangle\langle\rho_u^c\rho_u^c\rangle =
(N_c^2-1)^2 \,\frac{\mu^4}{a^4}\,.
\label{eq:rho4coefffactorizable}
\ee

On the other hand, in the lattice configuration where $x=y=u=v$,
the coefficient of the correlator is given by
\be
\langle\rho_{x}^a\rho_{x}^a\rho_{x}^c\rho_{x}^c\rangle_{\rm MV} =
\left[
(N_c^2-1)^2+2(N_c^2-1)
\right]
\frac{\mu^4}{a^4}
=(N_c^4-1)\,\frac{\mu^4}{a^4}\,,
\label{eq:color_factor_MV_connected_config}
\ee
having a different color factor from the factorizable case.

In the next section, we shall consider the Taylor expansion of the 
hypergeometric functions in two different regimes in order to 
study the limit of small as well as large non-Gaussian fluctuations. 
The results from these asymptotic cases will be compared to 
non-perturbative numerical calculations.

\subsection{The dilute regime: quartic term as small perturbation}\label{sec:IIIA}

In the $\kappa_4\to\infty$ limit the quartic term is a small perturbation.
Expanding \eq{eq:renormMuSU2} at $X\to\infty$ up to the order of $1/X$:
\be
\bar\mu^2\left( 1 - \frac{N_c^2+1}{4X} \right) = \mu^2\,.
\label{eq:PertResMuX}
\ee
Writing it in terms of $\kappa_4$ gives: 
\be
\mu^2 = \bar\mu^2 \left( 1 - 12\frac{\bar\mu^4}{\kappa_4}\frac{(N_{c}^2+1)}{\mathtt{a}^2} \right) = \bar\mu^2\,Z\,,
\label{eq:PertResMu}
\ee
which is the same\footnote{The factor $12$ in \eq{eq:PertResMu} is different from the 
factor $4$ present in Eq. (20) of Ref.~\cite{Dumitru:2011zz},
because the authors of Ref.~\cite{Dumitru:2011zz} changed
the definition of the coefficient of the quartic term
by a factor 1/3 ($3/\kappa_4\to 1/\kappa_4$) from \eq{eq:quarticAction}.} result 
obtained after a perturbative expansion of the quartic term in the 
small-$x$ action as done in Ref.~\cite{Dumitru:2011zz}.

According to~\cite{Dumitru:2011zz}, a contribution of order 
$1/\kappa_4^2$ renormalizes the $\bar\mu^8$ factor appearing 
in the correction to the four-point function of color charges 
at order $1/\kappa_4$. 
Expanding \eq{eq:rho4NonGauss} at $X\to\infty$ up to the term $1/X^2$ yields
\begin{align}
\langle \rho^a_{x}\rho^a_{x}\rho^c_{x}\rho^c_{x} \rangle_{\text{NG}}
&\simeq(N_c^4-1)
\frac{\bar\mu^4}{a^4}
\left[1 - \frac{N_c^2+2}{2X}+\frac{31+24N_c^2+5N_c^4}{16X^2}
\right] \\
&=(N_c^4-1)
\frac{\bar\mu^4}{a^4}
\left[1 - \frac{N_c^2+1}{2X}-\frac{1}{2X}\left(1-\frac{N_c^2+1}{X}\right)
   +\frac{23+16N_c^2+5N_c^4}{16X^2}
\right]
\,.
\label{eq:rho4Renorm}
\end{align}
We renormalize $\bar\mu^4$ and $\bar\mu^8$ 
by using \eq{eq:PertResMuX}:
\be
 \mu^4 \simeq \bar\mu^4\left(1-\frac{N_c^2+1}{2X}\right),~~~
 \mu^8 \simeq \bar\mu^8\left(1-\frac{N_c^2+1}{X}\right)\,, 
\ee
valid at order $1/\kappa_4$. The remaining terms of 
order $1/\kappa_4^2$ in \eq{eq:rho4Renorm} 
are discarded, as in~\cite{Dumitru:2011zz}.
Then, we obtain the correlator of four-color charges
in the dilute limit up to the order of $\kappa_4$~\cite{Dumitru:2011zz}:
\be
\langle \rho^a_{x}\rho^a_y \rho^c_u\rho^c_v \rangle_{\text{NG}} = (N_c^4-1)\,\mu^4
\bigg[\frac{\delta_{xy}}{a^2}\frac{\delta_{uv}}{a^2}
\left( 1 - 24\frac{\mu^4}{\kappa_4}\frac{\delta_{xu}}{a^2}\right)+
\frac{\delta_{xu}}{a^2}\frac{\delta_{yv}}{a^2}
\left( 1 - 24\frac{\mu^4}{\kappa_4}\frac{\delta_{xy}}{a^2}\right)
+\frac{\delta_{xv}}{a^2}\frac{\delta_{yu}}{a^2}
\left( 1 - 24\frac{\mu^4}{\kappa_4}\frac{\delta_{xy}}{a^2}\right)\bigg]\,.
\label{eq:4ptfunct_pert}
\ee
All other components are similar to this one, with the only difference 
being the permutation of the indexes of Kronecker's deltas. 
In the continuum notation, it reads
\bea
\langle \rho^a_{x}\rho^b_y \rho^c_u\rho^d_v \rangle_{\text{NG}} = &\mu^4&
\bigg[ 
\delta^{ab}\delta^{cd}
\delta(x-y)\delta(u-v)
\left( 1 - 24\frac{\mu^4}{\kappa_4}\delta(x-u)\right) 
+ 
\delta^{ac}\delta^{bd}
\delta(x-u)\delta(y-v)
\left( 1 - 24\frac{\mu^4}{\kappa_4}\delta(x-y)\right) \nn\\
&+& 
\delta^{ad}\delta^{bc}
\delta(x-v)\delta(y-u)
\left( 1 - 24\frac{\mu^4}{\kappa_4}\delta(x-y)\right) 
 \bigg]\,.
\label{eq:pertresultrho4}
\eea
The combination of (Dirac's) delta functions is such that the non-Gaussian
correction modifies the result from the MV model only if the four-point function of 
color charges is a local quantity, that is, $x=y=u=v$.  
On the other hand, in the configuration that each pair of color charges sit at different sites
({\it i}.{\it e}. $x=y$, $u=v$ but $x\neq u$), the four-point function factorizes into 
the product of two two-point functions,
and the result is identical to the one in the MV model. 

Since we are interested in the effect of the non-Gaussian correction 
to the MV model, we set $x=y=u=v$ in the lattice expression; 
in other words, we calculate \eq{eq:4ptfunct_pert} at the 
delta functions. 
The ratio of the correlator of four-color charges in the non-Gaussian 
to the Gaussian theory results in:  
\be
\frac{\langle \rho^a_{x}\rho^a_x \rho^c_x\rho^c_x \rangle_{\text{NG}}}{\langle \rho^a_{x}\rho^a_x \rho^c_x\rho^c_x \rangle_{\rm MV}} = 1 - 24\frac{\mu^4}{\kappa_4\,a^2}\,,
\label{eq:ratio4ptfunct_pert}
\ee
for the dilute limit.

\subsection{Large non-Gaussian fluctuations}

We now consider the limit of large non-Gaussian fluctuations, $Z\to0$,
where sizable deviations from the MV model are expected. 

Taylor expanding the left-hand side of \eq{eq:renormMuSU2} 
around $X\rightarrow 0$ up to the order $\mathcal{O}[X]$ yields: 
\be
\frac{\mu^2}{Z\,a^2}\left[
\frac{\Gamma \left(\frac{1}{4} \left(N_c^2+1\right)\right)}{\Gamma \left(\frac{1}{4} \left(N_c^2+3\right)\right)} \sqrt{X} + 
\left(\frac{ 2\Gamma \left(\frac{1}{4} \left(N_c^2+1\right)\right)^2}{\Gamma \left(\frac{1}{4} \left(N_c^2-1\right)\right)
\Gamma \left(\frac{1}{4} \left(N_c^2+3\right)\right)}-2\right)
X \right] = \frac{\mu^2}{\mathtt{a}^2} \nn
\ee
Thus, the expression up to the order of $Z$ is given by
\be
\frac{ \Gamma \left(\frac{1}{4} \left(N_c^2+1\right)\right)}{4\, \sqrt{3}\, \Gamma \left(\frac{1}{4} \left(N_c^2+3\right)\right)} \frac{\sqrt{\kappa_4}}{a}
+
\left(\frac{\Gamma \left(\frac{1}{4} \left(N_c^2+1\right)\right)^2}{\Gamma \left(\frac{1}{4} \left(N_c^2-1\right)\right) \Gamma \left(\frac{1}{4} \left(N_c^2+3\right)\right)}-1\right)
\frac{Z\,\kappa_{4}}{24\,\mu^2}= \frac{\mu^2}{\mathtt{a}^2}\,.
\label{eq:renormEqBeforeScenarios}
\ee
We will work out the solution of \eq{eq:renormEqBeforeScenarios}
for the three renormalization schemes in the next sections.

We now present an analytical expression for the local configuration 
$\langle\rho_x^a\rho_x^b\rho_x^c\rho_x^d\rangle$ in 
the regime of large non-Gaussian fluctuations. 
We begin by setting $x=y=u=v$, so that we calculate 
\eq{eq:coloravglattice} for this particular configuration.
For the color space, we have the following color contractions
$\delta^{ab}\delta^{cd} + \delta^{ac}\delta^{bd} + \delta^{ad}\delta^{bc}$.
It is sufficient to consider the case $a=b$ and $c=d$, 
as each term yields the same contribution.
For the large non-Gaussian fluctuation limit,
Taylor expanding \eq{eq:rho4NonGauss} around $X\rightarrow 0$
up to the order $\mathcal{O}[Z^2]$
yields: 
\bea
\langle \rho^a_{x}\rho^a_{x}\rho^c_{x}\rho^c_{x} \rangle_{\text{NG}} &=& 
(N_c^4-1)\, \bigg[
\frac{\kappa_4 }{12\, a^2 \left(N_c^2+1\right)}
-\frac{\sqrt{\pi}\, 2^{(7-N_c^2)/2}\, \Gamma \left(\frac{1}{2} \left(N_c^2+1\right)\right) }{3\,\sqrt{3}\, \left(N_c^2+1\right)\, \Gamma \left(\frac{1}{4} \left(N_c^2+3\right)\right)^2\,}
\frac{Z\,\kappa_4^{3/2}}{a\,\mu^2}\nn \\
&+& 
\frac{  \left(4 \, \Gamma \left(\frac{1}{4} \left(N_c^2+3\right)\right)^2-\left(N_c^2-1\right) \, \Gamma \left(\frac{1}{4} 
\left(N_c^2+1\right)\right)^2\right)}{1152 \left(N_c^2+1\right) \, \Gamma \left(\frac{1}{4} \left(N_c^2+3\right)\right)^2}\frac{Z^2\,\kappa_4^2}{\mu^4}
\bigg] 
\,.
\label{eq:rho4nonGaussNoScenario}
\eea

Let us now compute the four-point function of color charges
in the lowest order.
From 
\eq{eq:renormEqBeforeScenarios},
one obtains
\be
\kappa_4 = 48\,\frac{\Gamma \left(\frac{1}{4} \left(N_c^2+3\right)\right)^2}
     {\Gamma \left(\frac{1}{4} \left(N_c^2+1\right)\right)^2}\frac{\mu^4}{a^2}\,.
\label{eq:k4scenario1}
\ee
We see that $\kappa_4$ does not depend on $\bar\mu$.
For the four-point function of color charges in the leading order term in \eq{eq:rho4nonGaussNoScenario} results 
in
\be
\langle \rho^a_{x}\rho^a_{x}\rho^c_{x}\rho^c_{x} \rangle_{\text{NG}} = 
\frac{(N_c^4-1)\,\kappa_4}{12\, a^2\, \left(N_c^2+1\right)} =
\frac{4 \left(N_c^2-1\right)\, \Gamma \left(\frac{1}{4} \left((N_c^2+3\right)\right)^2}
     {\Gamma \left(\frac{1}{4} \left((N_c^2+1\right)\right)^2}\frac{\mu^4}{a^4} =
\frac{\left(N_c^4-1\right)\, \Gamma \left(\frac{1}{4} \left(N_c^2+3\right)\right)^2}{\Gamma \left(\frac{1}{4} \left(N_c^2+1\right)\right) \Gamma \left(\frac{1}{4} \left(N_c^2+5\right)\right)}\frac{\mu^4}{a^4}\,,
\label{eq:rho4_LO_scenario13}
\ee
where we used $\Gamma((N_c^2+1)/4) = 4\,\Gamma((N_c^2+5)/4)\,/\,(N_c^2+1)$.

The correlator of four-color charges in the non-Gaussian theory follows
the same structure as~\eq{eq:pertresultrho4} in continuum notation:
\bea
\langle \rho^a_{x}\rho^b_{y} \rho^c_{u}\rho^d_{v} \rangle_{\text{NG}} &=&  
\mu ^4
\bigg\{ 
\delta^{ab}\delta^{cd}\delta(x-y)\delta(u-v)\left[1 - {\mathcal{C}_\mathrm{NG}}\frac{\mu^4}{\kappa_4}\,\delta(x-u) \right] + 
\delta^{ac}\delta^{bd}\delta(x-u)\delta(y-v)\left[1 - {\mathcal{C}_\mathrm{NG}}\frac{\mu^4}{\kappa_4}\,\delta(x-y) \right] \nonumber \\ 
&+& \delta^{ad}\delta^{bc}\delta(x-v)\delta(y-u)\left[1 - {\mathcal{C}_\mathrm{NG}}\frac{\mu^4}{\kappa_4}\,\delta(x-y) \right]
\bigg\}\,,
\label{eq:LeadingOrderRho4NGSU2continuum}
\eea
where ${\mathcal{C}_\mathrm{NG}}$ reads
\be
{\mathcal{C}_\mathrm{NG}} = 48\frac{\Gamma(\frac{1}{4}(N_c^2+3))^2}{\Gamma(\frac{1}{4}(N_c^2+1))^2} 
\bigg[ 1 -  \frac{\Gamma \left(\frac{1}{4} \left(N_c^2+3\right)\right)^2}{\Gamma \left(\frac{1}{4} \left(N_c^2+1\right)\right) \Gamma \left(\frac{1}{4} \left(N_c^2+5\right)\right)} \bigg]\,.
\label{eq:nonGaussCorrection}
\ee

As in the MV model, the color factor multiplying $\mu^4/{\mathtt{a}^4}$ 
depends on the spatial configuration in which the correlator is calculated: 
\bea
\langle \rho^a_{x}\rho^a_{y} \rho^c_{u}\rho^c_{v}\rangle_{\text{NG}} \propto \,\,\,
\begin{cases}
	(N_c^2-1)^2\,, & \text{if } x = y,\,\, u=v\,\, (u\ne x) \label{eq:nongauss_propfactor1}\\
	(N_c^4-1)\,
	\frac{\Gamma \left(\frac{1}{4} \left(N_c^2+3\right)\right)^2}{\Gamma \left(\frac{1}{4} \left(N_c^2+1\right)\right) \Gamma \left(\frac{1}{4} \left(N_c^2+5\right)\right)}\,,  & \text{if }  x=y=u=v \label{eq:nongauss_propfactor2} \,.
\end{cases}
\eea

We then consider the ratio of the correlator of four-color charges 
from the non-Gaussian to the Gaussian theories for the configurations shown above. 
When setting $x = y$, $u=v$ with the condition $u\ne x$ the non-Gaussian 
correction is not present, and we have 
\be
\frac{\langle \rho^a_{x}\rho^a_{x} \rho^c_{u}\rho^c_{u} \rangle_{\text{NG}}}{\langle \rho^a_{x}\rho^a_{x} \rho^c_{u}\rho^c_{u} \rangle_{\text {MV}}} = 
\frac{\langle \rho^a_{x}\rho^a_{x}\rangle \langle \rho^c_{u}\rho^c_{u} \rangle}{\langle \rho^a_{x}\rho^a_{x} \rangle \langle \rho^c_{u}\rho^c_{u} \rangle } = 
\frac{(N_c^2-1)^2\,\mu^4 / a^4}{(N_c^2-1)^2\,\mu^4 / a^4} = 1\,,
\label{eq:RatioRho22NGSUNc}
\ee
as expected (see \eq{eq:rho4coefffactorizable}). On the other hand, 
for the configuration where
$x=y=u=v$, the ratio of  $\langle \rho^a_{x}\rho^a_{x} \rho^c_{x}\rho^c_{x} \rangle$ 
from the non-Gaussian theory to the Gaussian theory yields: 
\be
\frac{\langle \rho^a_{x}\rho^a_{x} \rho^c_{x}\rho^c_{x} \rangle_{\text{NG}}}{\langle \rho^a_{x}\rho^a_{x} \rho^c_{x}\rho^c_{x} \rangle_{\text {MV}}} = 
\frac{\Gamma \left(\frac{1}{4} \left(N_c^2+3\right)\right)^2}{\Gamma
\left(\frac{1}{4} \left(N_c^2+1\right)\right) \Gamma \left(\frac{1}{4}
\left(N_c^2+5\right)\right)}\,,
\label{eq:RatioRho4NGSUNc}
\ee
showing that, in the $Z\rightarrow 0$ limit,
the ratio of correlators of color charge depends 
only on the number of colors $N_c$ (so it is 
constant for fixed $N_c$). 
For SU(2) and SU(3), 
\eq{eq:RatioRho4NGSUNc} evaluates to 
\begin{subequations}
\begin{empheq}[left={
{\dfrac{\langle \rho^a_{x}\rho^a_{x} \rho^c_{x}\rho^c_{x}\rangle_{\text{NG}}}
{\langle \rho^a_{x}\rho^a_{x} \rho^c_{x}\rho^c_{x} \rangle_{\text{MV}}} =}
\empheqlbrace}]{alignat=2}
0.822504  & \quad \text{for SU(2)} \label{eq:nongauss_effect_SU2}\\
0.905415  & \quad \text{for SU(3)} \label{eq:nongauss_effect_SU3}\,,
\end{empheq}
\end{subequations}
so the correlator of four-color charges calculated at the 
delta functions in a lattice setup should decrease 
by $\sim 18\%$ ($\sim 10\%$) in the non-Gaussian theory 
compared to the Gaussian theory for SU(2) (SU(3)) 
in the limit of very large non-Gaussian fluctuations. 

Finally, we note that there exist two different conditions 
where one may factorize four-point functions (and other 
higher-order correlators) of color charges into products of 
two-point functions~\cite{Dumitru:2010mv}: when using a Gaussian weight 
function for color average, as the MV model, and the large-$N_c$ limit. 
For the large-$N_c$ limit in our case, the ratio \eq{eq:RatioRho4NGSUNc}
evaluates to one:
\be
\lim\limits_{N_c\rightarrow \infty}
\frac{\langle \rho^a_{x}\rho^a_{x} \rho^c_{x}\rho^c_{x}\rangle_{\text{NG}}}
{\langle \rho^a_{x}\rho^a_{x} \rho^c_{x}\rho^c_{x} \rangle_{\text{MV}}}
=
\lim\limits_{N_c\rightarrow \infty}	
 \frac{\Gamma \left(\frac{1}{4} \left(N_c^2+3\right)\right)^2}{\Gamma
 \left(\frac{1}{4} \left(N_c^2+1\right)\right) \Gamma \left(\frac{1}{4}
 \left(N_c^2+5\right)\right)} = 1 \,.
\ee

\section{Renormalization schemes}

In this section, 
we consider two opposite perturbative regimes, 
small and large non-Gaussian fluctuations,
in three different renormalization schemes.
These results are then compared to a full 
non-perturbative calculation, where 
\eq{eq:renormMuSU2} is solved numerically 
for SU(2) and SU(3) color symmetry groups.
We assume a constant average color 
charge within the nuclear system and invoke 
the infinite nucleus approximation.
Note that the infinite nucleus approximation 
does not imply an infinite number of color charges 
in lattice calculations. Therefore, one can still 
study deviations from a Gaussian ensemble 
even in this simplified scenario.

\subsection{Multi-point correlators of color charges and the renormalization equation in SU($N_c$) for 
	the first renormalization scheme}

The first renormalization scheme is defined by 
the condition:
\be
\frac{\kappa_{4}}{\bar\mu^6} \equiv \lambda = \frac{\gamma}{g^2}\,,
\label{eq:scheme1}
\ee
where $\gamma= 48\, (144)$ for SU(2) (SU(3)), 
motivated by \eq{eq:nongausscouplingSU3} and \eq{eq:nongausscouplingSU2}.

In the limit of small non-Gaussian fluctuations,
using \eq{eq:scheme1} in \eq{eq:PertResMu}  
and rewriting it in terms of $Z(a) = \mu^2/\bar\mu^2(a)$,
\be
Z(a)
=\frac{\mu^2 a^2}{12(N_{c}^2 + 1)\,g^2/\gamma + \mu^2\, a^2}\,
\label{eq:Z1} \,,
\ee
shows that the renormalization factor decreases with the lattice spacing 
and the perturbative calculation will break down at some point for small $a$.
The condition for small non-Gaussian fluctuations, $Z\approx 1$,
requires $\mu^2 a^2\, \gg\, 12(N_c^2+1)\, g^2/\gamma$.
We also note that $0 < Z \le 1$.


Using \eq{eq:scheme1} and \eq{eq:Z1} in \eq{eq:ratio4ptfunct_pert}, 
one can write the ratio of the four-point function of color 
charge in the non-Gaussian theory to that in the MV model as
\be
\frac{\langle \rho^a_{x}\rho^a_x \rho^c_x\rho^c_x \rangle_{\text{NG}}}
{\langle \rho^a_{x}\rho^a_x \rho^c_x\rho^c_x \rangle_{\text{MV}}} =
1 - \frac{24\,(g^2/\gamma)\,\mu^4\,a^4}{\left[\,  \mu^2\,a^2 + 12\,(N_c^2+1)\,g^2/\gamma\,\right]^3}\,. 
\label{eq:Rho4scenario1pert}
\ee

In this renormalization scheme, non-Gaussian fluctuations increase
with the lattice spacing since $Z\to 0$ as $a\to0$.
Thus, one cannot discuss the continuum limit within the 
perturbative calculation in the limit of small non-Gaussian 
fluctuations in this renormalization scheme\footnote{One could 
	consider the $a\rightarrow a_0 \, (a_{0}>0)$ 
	limit, thus attributing a physical meaning 
	to the lattice spacing: the definition of the 
	ultraviolet cutoff in (transverse) momentum space, 
	$p_\perp^{\text max}= \pi /a_0$. This  case
	implies that theories with different cutoffs 
	will produce different results for the correlators 
	sensitive to the non-Gaussian correction to the MV 
	model~\cite{Dumitru_priv_comm}. In this work 
	we only consider the $a\rightarrow 0$ limit.}.

In the limit of large non-Gaussian fluctuations,
using \eq{eq:scheme1} in \eq{eq:renormEqBeforeScenarios}
yields
\be
\frac{ \Gamma \left(\frac{1}{4} \left(N_c^2+1\right)\right)}{4\, \sqrt{3}\, \Gamma \left(\frac{1}{4} \left(N_c^2+3\right)\right)} \frac{\sqrt{\gamma}\,\mu^3}{g\,a\,Z^{3/2}}
+
\left(\frac{\Gamma \left(\frac{1}{4} \left(N_c^2+1\right)\right)^2}{\Gamma \left(\frac{1}{4} \left(N_c^2-1\right)\right) \Gamma \left(\frac{1}{4} \left(N_c^2+3\right)\right)}-1\right)
\frac{\gamma \,\mu^4}{24\,g^2\,Z^2}= \frac{\mu^2}{\mathtt{a}^2}\,.
\label{eq:renormEqScenario1}
\ee
\eq{eq:renormEqScenario1} can be written as a quartic equation for $Z$, 
thus it can be solved. At the leading order,
$Z^3$ is proportional to the square of the lattice spacing:
\be
\frac{ \Gamma \left(\frac{1}{4} \left(N_c^2+1\right)\right)}{4\, \sqrt{3}\, \Gamma \left(\frac{1}{4} \left(N_c^2+3\right)\right)} \frac{\sqrt{\gamma}\,\mu^3}{g\,a\,Z^{3/2}}
= \frac{\mu^2}{\mathtt{a}^2}\,
\quad\rightarrow\quad
Z^3 = 
\frac{ \gamma\, \Gamma \left(\frac{1}{4} \left(N_c^2+1\right)\right)^2}{48\, g^2\, \Gamma \left(\frac{1}{4} \left(N_c^2+3\right)\right)^2}\,a^2\, \mu^2 \,.
\label{eq:renormEqScenario1_cont}
\ee
Therefore, any system is substantially affected by non-Gaussian 
corrections in the continuum limit.
\begin{figure}[htb]
	\begin{center}
		\includegraphics[scale=0.68]{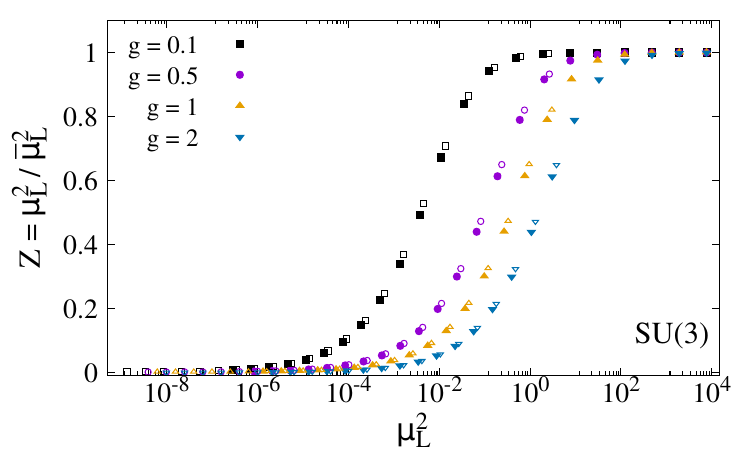}
		\includegraphics[scale=0.68]{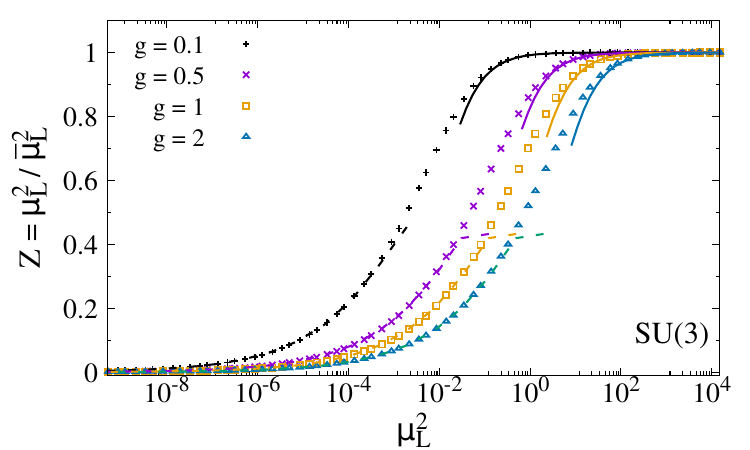}\\
	\end{center}
	\vspace*{-7mm}
	\caption[a]{(Left) Lattice spacing dependence of the renormalization 
		factor $Z$ from a non-perturbative calculation for SU(3) 
		for (filled points) $L=11.5$ fm and $\mu=3$ GeV and 
		(open points) $L\approx 1.77$ fm and $\mu=0.35$ GeV showing scaling invariance.
		(Right) Comparison of analytical results with the non-perturbative 
		calculation. Solid lines are the result from \eq{eq:Z1}, valid for $Z\sim 1$; 
		dashed lines represent the solution of \eq{eq:renormEqScenario1}, 
		valid for $Z\sim 0$, with the leading and next-to-leading order 
		terms in lattice spacing. For asymptotically small values of 
		$\mu_L$ the dashed curves reduce to \eq{eq:renormEqScenario1_cont}.
		}
	\label{fig:RenormParamSU3}
\end{figure}

Next we consider the lattice spacing dependence of the 
renormalization factor $Z$. Following~\cite{Krasnitz:1998ns}, 
we write $\mu_L=\mu\,a$. 
The filled points in left panel of~\fig{fig:RenormParamSU3} 
were obtained for a lattice of size $L = 11.5$ fm, 
which corresponds to the radius $R=6.5$ fm of 
a gold nucleus by the relation $L^2=\pi R^2$, and 
using $\mu = 3$ GeV in the MV model. 
As $\mu$ is kept fixed, the $\mu_L$ dependence is obtained 
by solving \eq{eq:renormalizationMu} for decreasing values of the lattice 
spacing, which are obtained by successively increasing the number of 
sites of the lattice by a factor of two while keeping its volume fixed
($L^2 = N_s^2\, a^2=$ constant) at each step. 
For instance, the rightmost point is the result for a lattice with 
$N_s = 2$, the next one is the result for a lattice with $N_s = 2^2$ and so on, 
with the last point shown in the figure corresponding to a lattice with $N_s = 2^{22}$. 
As the infinite nucleus approximation throws away any detailed information
about the geometry of all physical systems, one should expect exact scale invariance.
This means that the only difference between a hadron and a heavy nucleus should be 
the size of the lattice in physical units, so both are related by a simple scaling factor. 
Consequently, once the coupling $g$ is fixed, results for different systems should all fall 
under the same curve, with all physics being controlled by the dimensionless quantity $\mu_L$.
To show that this is the case, the left panel of \fig{fig:RenormParamSU3} 
also includes the results (open symbols) for a system with 
$L = 1\, \mathrm{fm}\, \sqrt{\pi} \approx 1.77$ fm, 
with $\mu = 0.35$ GeV (which loosely corresponds to a proton). 
One clearly sees that the results for both systems fall 
under the same curve, showing the scaling invariance, as expected. 
Therefore, it is only needed to specify the details of a given 
system (here completely determined by the lattice size in 
physical units and color charge $\mu$) when discussing results 
at fixed lattice size.

In the right panel of \fig{fig:RenormParamSU3}, we compare 
the resulting lattice spacing dependence  
of the renormalization factor $Z(a)$ from the asymptotic 
cases considered above with the result from a full
non-perturbative calculation for different couplings.
We see that $Z\rightarrow 0$ as 
$a\rightarrow 0$ for all values of the coupling 
in the full non-perturbative calculation, 
indicating that \eq{eq:quarticAction}
``flows" towards a theory dominated 
by large non-Gaussian fluctuations. 
In this renormalization scheme, even though $\bar\mu(a)$ has been determined by requiring 
the matching of the two-point function of color charges 
in the non-Gaussian theory and the MV model for each lattice 
size, the matching is achieved by decreasing the renormalization 
factor, that is, by moving further away from the MV model regardless 
of the system size.
On purely theoretical grounds, nothing is prohibiting such weight 
functions to exist, however, such a scenario seems unlikely to 
be realized. 

\begin{figure}[htb]
	\begin{center}
		\includegraphics[scale=0.68]{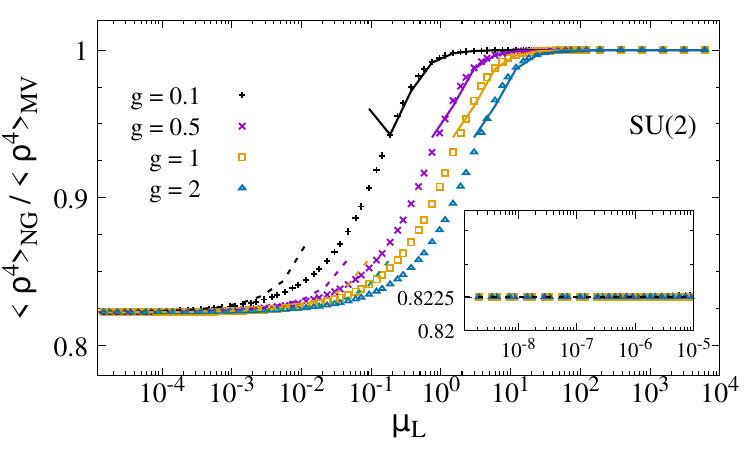}
		\includegraphics[scale=0.68]{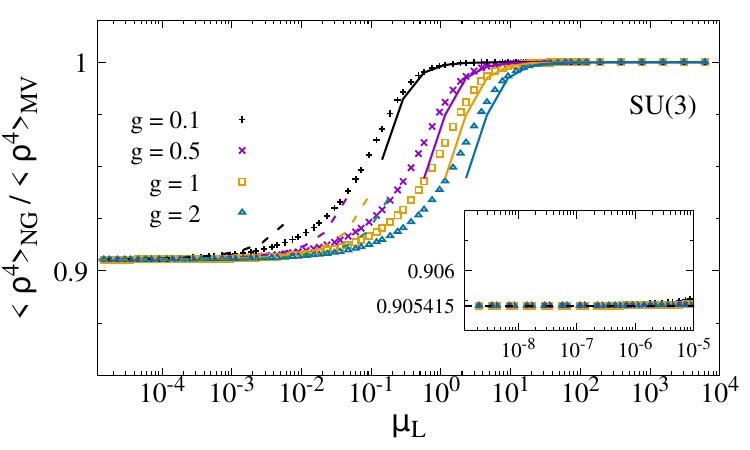}
	\end{center}
	\vspace*{-7mm}
	\caption[a]{The ratio of the correlator of four-color charges in the non-Gaussian to the 
		Gaussian ensemble as a function of $\mu_L$ for (left) SU(2) and (right) SU(3) 
		color symmetry group. 
		The solid lines represent the perturbative results for $Z\sim 1$, 
		\eq{eq:Rho4scenario1pert}, while the dashed ones are the result 
		of solving~\eq{eq:renormEqScenario1} with all terms presented. 
		For asymptotically small values $\mu_L$, which corresponds to
		the $Z\to0$ limit,
		the dashed curves reduce to the values
		in \eq{eq:nongauss_effect_SU2} for SU(2)
		and \eq{eq:nongauss_effect_SU3} for SU(3),
		respectively. The inset plots extend our results up to $\mu_L\sim 10^{-8}$, 
		showing that indeed the results converged to the continuum limit.}
	\label{fig:ratioLambdasVSlatspacing}
\end{figure}

\fig{fig:ratioLambdasVSlatspacing} shows the lattice spacing 
dependence of the ratio of the correlator of four-color charges
in the non-Gaussian ensemble to the Gaussian ensemble 
for different values of the coupling.
As $\bar\mu\neq \mu$, results from the Gaussian and non-Gaussian ensembles 
would fall in different bins in the horizontal axis, and a comparison between 
them would only be possible after extrapolating the results to the continuum 
limit. This is circumvented by using the correlator of two-color charges to 
form  dimensionless quantities: $a^4\langle\rho^2\rangle = a^2\mu^2 = \mu_L^2$. 
This is equivalent to assuming the average color charge from the MV model
as a momentum scale in the horizontal axis.
Our results show that
i) for $\mu_L \gg g$ there is no deviation from the Gaussian theory,
and the ratio is one; 
ii) for $\mu_L \lesssim g$ there is a smooth transition from a 
Gaussian dominated distribution (where the perturbative calculation 
from~\cite{Dumitru:2011zz} applies) 
to a distribution which is more and more dominated by the quartic term. 
The resulting effect is the gradual reduction of the higher-order correlator of 
color charges, in accordance with increasing deviations from the MV
model presented \fig{fig:RenormParamSU3};
iii) such transition shows a hierarchy with the coupling constant, the agreement 
with the perturbative result breaks first for larger values of $g$ at fixed $\mu_L$; 
iv) once the distribution of color charges is dominated by the non-Gaussian 
term ($Z\to 0$), the ratio converges to the continuum limit value shown in 
\eq{eq:nongauss_effect_SU2} for SU(2)
and \eq{eq:nongauss_effect_SU3} for SU(3) for all values of the 
coupling and $\mu_L$. 

\begin{figure}[htb]
	\begin{center}
		\includegraphics[scale=0.40,trim={0 0.8cm 0 0},clip]{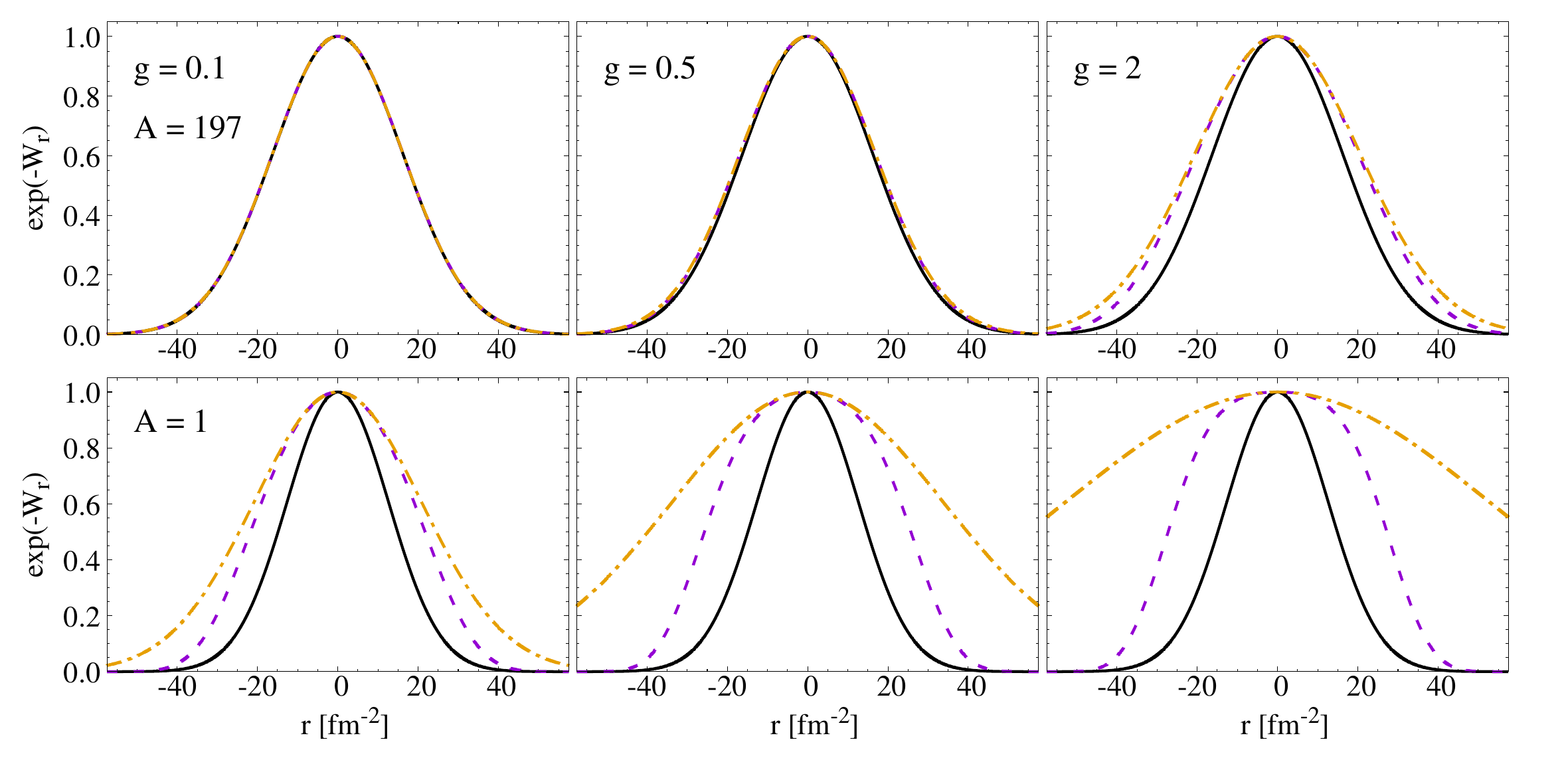}
	\end{center}
	\vspace*{-7mm}
	\caption[a]{Weight function from a Gaussian (solid line) and 
		non-Gaussian (dashed line) ensembles for a system with 
		(top panels) $L = 11.5$ fm and $\mu = 3.0$ GeV and
		(bottom panels) $L \approx 1.77$ fm and $\mu = 0.35$ GeV
		and different values of the coupling in SU(3) in a lattice 
		with $N_s=128$. 
		The dashed-dotted line is a Gaussian distribution with
		the standard deviation equal to the renormalized average 
		color charge $\bar\mu$. }
	\label{fig:ProbDistFuncts}
\end{figure}

Let us look at how far away the non-Gaussian distribution
is from a Gaussian distribution. 
\fig{fig:ProbDistFuncts} shows the weight function from the MV model (full line) 
and the respective non-Gaussian ensemble (dashed line) 
for (top panels) a Gold-like system ($L = 11.5$ fm) with $\mu=3$ GeV 
and (bottom panels) a proton-like system
($L=1\,\mathrm{fm}\,\sqrt{\pi}\approx 1.77$ fm) with $\mu = 0.35$ GeV
for different values of $g$ for SU(3) 
as a function of $r^2\equiv \sum_{a=1}^{N_c^2-1}(\rho^{a}_{x})^{2}$. 
These weight functions were obtained in a lattice with $N_s = 128$,  
corresponding to $\mu_L\sim 1.37$ ($\mu_L\sim 0.025$) 
for a Gold-like (proton-like) system. 
The dashed-dotted line represents a Gaussian distribution
with the standard deviation $\bar\mu_L$.
We note that the distributions for SU(2) have the same features.
The color charge distribution in the large system with small $g$
is described well as Gaussian.
For $g=2$, the quartic term starts to dominate,
and the resulting distribution gradually deviates from a 
Gaussian distribution. On the other hand, small systems already 
present strong deviations from the perturbative regime for all 
values of $g$ considered.

\subsection{Renormalization equation in SU($N_c$) for the 
second renormalization scheme}\label{sec:Scenario2}

In this section, we consider the second renormalization scheme, in which
$\kappa_{4}$ is kept fixed to a given constant value. This renormalization 
presents an important difference from the previous one: 
$\kappa_4$ and $Z$ are treated as independent parameters.

In the regime where the quartic term is a small perturbation,
$Z$ is determined by $\kappa_4$ through \eq{eq:PertResMu}: 
\be
Z(a,\kappa_4) = 1 - \frac{12\,(N_c^2+1)\, \bar\mu^4}{\kappa_4\, a^2}\,.
\label{eq:RenormalizedZfactorScenario2}
\ee
As we keep the calculation at order ${\mathcal{O}}[1/\kappa_{4}]$, 
we replace\footnote{At the level of perturbation theory, the contribution  
$32\cdot9\, (N_c^2+1)^2 \,\frac{\bar\mu^8}{\kappa^2_4 \,a^4}$, 
which involves a term of order $1/\kappa_4^2$, 
induces the shift $\bar\mu^4/\kappa_4 \to \mu^4/\kappa_4$.} 
$\bar\mu^4/\kappa_4$ by $\mu^4/\kappa_4$,
and $Z$ is given by:
\be
Z(a,\kappa_4) = 1 - \frac{12\,(N_c^2+1)\, \mu^4}{\kappa_4\, a^2}\,.
\label{eq:RenormalizedZfactorScenario2}
\ee

On the other hand, in the limit of large non-Gaussian fluctuations,
solving \eq{eq:renormEqBeforeScenarios} for $Z$ yields: 
\be
Z(a,\kappa_{4}) = 
\frac{2\, \mu^2\, \Gamma \left(\frac{1}{4} \left(N_c^2-1\right)\right) \left[ \sqrt{3\,\kappa_{4}}\, a\, \, \Gamma \left(\frac{1}{4} \left(N_c^2+1\right)\right)-12\, \mu^2\, \Gamma \left(\frac{1}{4} \left(N_c^2+3\right)\right)\right]}{\left[ 
\Gamma \left(\frac{1}{4} \left(N_c^2-1\right)\right) \Gamma \left(\frac{1}{4} \left(N_c^2+3\right)\right)
-\Gamma \left(\frac{1}{4} \left(N_c^2+1\right)\right)^2
\right]\,a^2\, \kappa_4}\,.
\label{eq:Zscenario2}
\ee
\eq{eq:Zscenario2} fixes $Z(a,\kappa_4)$ so that the 
non-Gaussian action reproduces the two-point function of 
color charges from the Gaussian theory. 
We note that $Z(a,\kappa_{4})$ will change the sign 
at some point in this second renormalization scheme. 
The sign of $Z(a,\kappa_{4})$ is determined by the factor 
$\sqrt{3\,\kappa_{4}}\, a\, \, \Gamma \left(\frac{1}{4} 
\left(N_c^2+1\right)\right)-12\, \mu^2\, \Gamma 
\left(\frac{1}{4} \left(N_c^2+3\right)\right)$.  
In particular, $Z$ will become zero when 
\be
a = \frac{4\,\sqrt{3}\, \Gamma\left(\frac{1}{4}
\left(N_c^2+3\right)\right)}{\Gamma \left(\frac{1}{4} 
\left(N_c^2+1\right)\right)}
\frac{\mu^2}{\sqrt{\kappa_{4}}}\,,
\label{eq:scenario2_z0}
\ee
and the renormalization factor becomes negative for
the lattice spacing smaller than the value given by \eq{eq:scenario2_z0}.
In fact, as $a\rightarrow 0$, \eq{eq:Zscenario2} 
becomes 
\be
Z(a,\kappa_{4}) = -\frac{24\, \mu^4}{\left(1-\frac{\Gamma \left(\frac{1}{4} \left(N_c^2+1\right)\right)^2}{\Gamma \left(\frac{1}{4} \left(N_c^2-1\right)\right) \Gamma \left(\frac{1}{4} \left(N_c^2+3\right)\right)}\right)\,a^2 \kappa_4}\,=\, -\frac{\mu^4}{a^2 \, \kappa_{4}} \times
\begin{cases}
	88.718,  & \text{for } \mathrm{SU(2)}\\
	206.138,  & \text{for } \mathrm{SU(3)}\,,
\end{cases}
\ee
thus, one cannot take the continuum limit in this renormalization scheme.

\begin{figure}[htb]
	\begin{center}
		\includegraphics[scale=0.68,trim=0.0 0.0 0.0 0.9cm,clip] {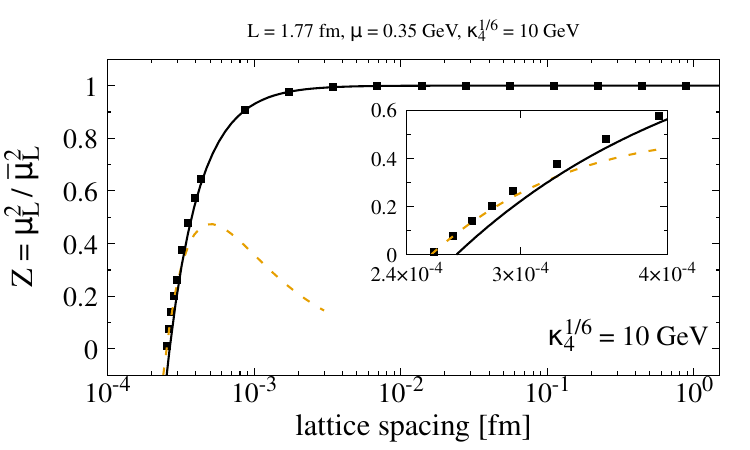}
		\includegraphics[scale=0.68,trim=0.0 0.0 0.0 0.9cm,clip] {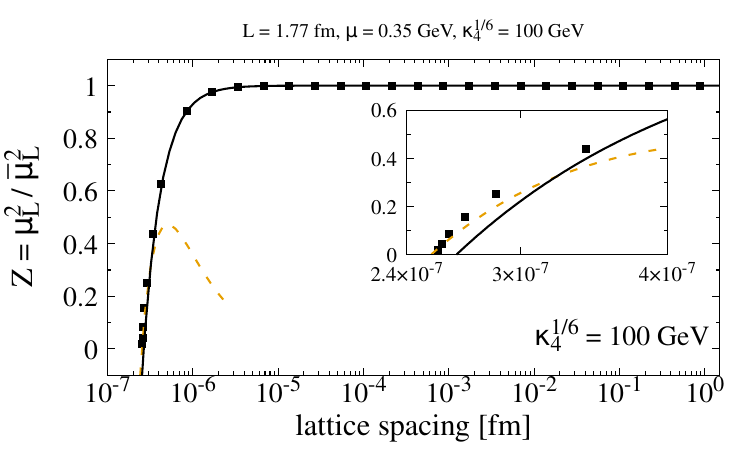}
	\end{center}
	\vspace*{-7mm}
	\caption[a]{Renormalization factor $Z(a,\kappa_4)$ in the second 
		renormalization scheme in SU(3) from full non-perturbative 
		calculation (points) for (left) $\kappa_{4}^{1/6}=10$ GeV and (right) 
		$\kappa_{4}^{1/6}=100$ GeV. The solid curve is the result from 
		\eq{eq:RenormalizedZfactorScenario2}. 
		The dashed curve is the result from \eq{eq:Zscenario2}.	}
	\label{fig:Scenario2_RenormFactor}
\end{figure}

\fig{fig:Scenario2_RenormFactor} shows the lattice spacing dependence 
of the renormalization factor in this renormalization scheme. 
The points represent the result from a numerical calculation in SU(3)
for a proton-like system ($\mu= 0.35$ GeV and $L\approx 1.77$ fm) 
for $\kappa_{4}^{1/6}= 10$ GeV and $\kappa_{4}^{1/6}= 100$ GeV. 
We expect similar behavior for other parameters.
The solid curve in each panel represents the result from 
\eq{eq:RenormalizedZfactorScenario2} valid for $Z\sim1$.
The dashed curve is the result from \eq{eq:Zscenario2} valid for $Z\sim 0$. 
\eq{eq:RenormalizedZfactorScenario2}
is in accordance with the non-perturbative calculation 
for $Z\sim 1$,
while \eq{eq:Zscenario2}
is able to match the non-perturbative calculation for $Z\to 0$. 

As in the previous renormalization scheme, 
the renormalization factor $Z(a,\kappa_4)$ 
is close to one at large $a$, indicating no 
deviation from the Gaussian theory.
However, its dependence with the lattice spacing changes quite 
drastically as $a\to 0$, with $Z(a,\kappa_{4})$ now presenting 
a sharper decrease. 
\eq{eq:Zscenario2} matches the full numerical calculation 
in the $Z\rightarrow 0$ limit. 
We verified via a numerical calculation
that once $Z=0$, there is no solution for the 
renormalization equation as
we only have the quartic term, 
whose coupling $\kappa_{4}$ is kept fixed
in this renormalization scheme.

\subsection{Multi-point correlators of color charges and
	the renormalization equation in SU($N_c$) for the third 
	renormalization scheme}\label{sec:Scenario3}

In the third renormalization scheme,
the renormalization factor $Z$ is kept fixed.
Thus, this renormalization scheme provides 
a way to control deviations from the MV model even in 
the limit of large non-Gaussian fluctuations. 

In the regime of small non-Gaussian fluctuations,
from \eq{eq:RenormalizedZfactorScenario2},
$\kappa_4$ is given by
\be
\kappa_4 (a,Z) = 12\frac{\mu^4}{1 - Z}\frac{(N_{c}^2+1)}{a^2}\,.
\label{eq:PertResMuScenario3}
\ee
The $Z\rightarrow 1$ limit leads to  $\kappa_{4}\rightarrow \infty$, 
thus recovering the MV model. The $a\rightarrow 0$ limit 
with $Z < 1$ also leads to $\kappa_{4}\rightarrow \infty$.
However, this does not mean it reduces exactly to the MV model
as far as  $Z$ is different from one.

The four-point function of color charges is obtained by
substituting $\kappa_{4}$ from \eq{eq:PertResMuScenario3} into
\eq{eq:ratio4ptfunct_pert},
\be
\frac{\langle \rho^a_{x}\rho^a_x \rho^c_x\rho^c_x \rangle_{\text{NG}}}
{\langle \rho^a_{x}\rho^a_x \rho^c_x\rho^c_x \rangle_{\text{MV}}} 
= 1 - \frac{2\,(1-Z)}{(N_{c}^2+1)}\,.
\label{eq:scenario1pertres}
\ee
We see that the ratio 
of four-point functions in the non-Gaussian to the 
Gaussian theory is independent of the lattice spacing.
This is a exclusive feature of this renormalization scheme,  
given that $Z$ changes with the lattice spacing in
the other two renormalization schemes.

In the limit of large non-Gaussian fluctuations,
the renormalization equation (\eq{eq:renormEqBeforeScenarios}) is
a quadratic equation for $\kappa_4$, and
has two solutions:
\be
\kappa_{4}(a,Z)\equiv \frac{\kappa_{4}(Z)}{a^2} = 
\frac{6\, \mu^4\,\left[ \alpha(Z) \pm \beta(Z) \right] }{a^2\, Z^2\, \Gamma_{\frac{1}{4},1}^2\, 
( \Gamma_{\frac{1}{4},1}^2 - \Gamma_{\frac{1}{4},-1}\, \Gamma_{\frac{1}{4},3})^2}
\label{eq:kappa4Scenario3}
\ee
where $\Gamma_{k,m}\equiv\Gamma \left( k\,(N_c^2 + m)\right)$ and 
\bea
\alpha(Z) &=& 2^{3-N_c^2}\,\pi\, \Gamma_{\frac{1}{2},-1}^2 \left( \, \Gamma_{\frac{1}{4},1}^2 \left(\left(N_c^2-1\right) Z+1\right)-4\, Z\, \Gamma_{\frac{1}{4},3}^2\,\right)\\
\beta(Z) &=& \Gamma_{\frac{1}{4},-1}^{3/2}\, \Gamma_{\frac{1}{4},1}^3 \, 
  \left[ 8\, Z\, \Gamma_{\frac{1}{4},3}\, \Gamma_{\frac{1}{4},1}^2+\Gamma_{\frac{1}{4},-1} 
 \, \left(\Gamma_{\frac{1}{4},1}^2-8 \, Z\, \Gamma_{\frac{1}{4},3}^2\right)\right]^{1/2}\,.
\eea
We verified that the two solutions above lead to different 
results in the $Z\to0$ limit. 
Setting $N_c=3$ in order to have a compact expression and further 
expanding the solution above proportional to  $\alpha(Z) - \beta(Z)$ 
around $Z=0$, then dividing it by the leading order expression 
for $\kappa_{4}$ (\eq{eq:k4scenario1}) gives 
\be
\frac{\kappa_{4}^{\text{LO+NLO}}} {\kappa_{4}^{\text{LO}}} = 1 + 1.05415\, Z + 1.38903\, Z^2\,.
\ee
Repeating the same procedure with the solution proportional to 
$\alpha(Z) + \beta(Z)$ yields terms proportional to 
$1/Z^2$ and $1/Z$, thus not recovering the leading order 
solution. Because of this, we discard such a solution.

Let us turn now to the computation of the four-point 
function of color charges. 
Using the solution for $\kappa_{4}$ proportional to 
$\alpha(Z)-\beta(Z)$ in \eq{eq:rho4nonGaussNoScenario} 
provides an expression for the correlator of four-color 
charges at the delta functions for an arbitrary value of 
$N_c$. 
The ratio of the correlator of color charges 
in the non-Gaussian to the Gaussian theory at 
leading order in the renormalization factor 
can be written as: 
\bea
\frac{\langle \rho^a_{x}\rho^a_{x} \rho^c_{x}\rho^c_{x} \rangle_{\text{NG}}}{\langle \rho^a_{x}\rho^a_{x} \rho^c_{x}\rho^c_{x} \rangle_{\text {MV}}} =
&&\frac{(-d_0+D)^2}{4(N_c^2+1)\,d_1^2\,Z^2}
\left(1+\frac{d_0}{d_1}(d_0 - D)\right)\,,\qquad
D=\sqrt{d_0^2 + 2d_1\,Z}
\label{eq:scenario3nonpert}
\eea
where 
\bea
d_0 &\equiv& \frac{1}{4}\frac{\Gamma((N_c^2+1)/4)}{\Gamma((N_c^2+3)/4)}\,,\\
d_1 &\equiv& d_0^2\,(N_c^2 -1)-\frac{1}{4}\,.
\eea
As in~\eq{eq:scenario1pertres}, 
valid in the limit of small non-Gaussian 
fluctuations, this ratio remains independent 
of the lattice spacing and is constant for  
fixed $Z$ in the limit of large 
non-Gaussian fluctuations.

\begin{figure}[htb]
	\begin{center}
		\includegraphics[scale=0.7]{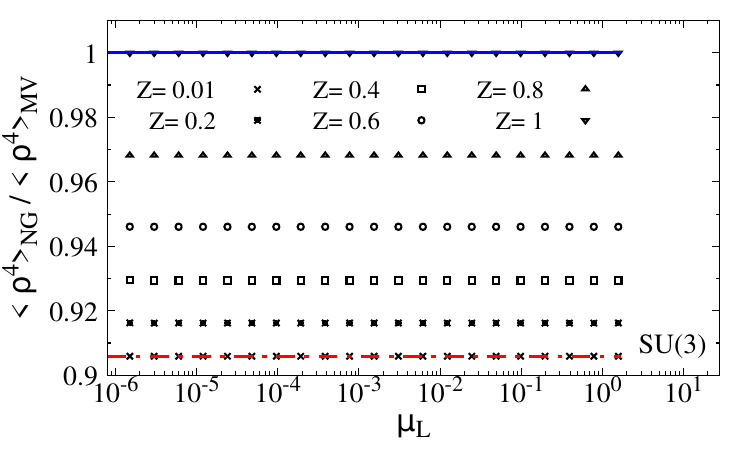}
		\includegraphics[scale=0.7]{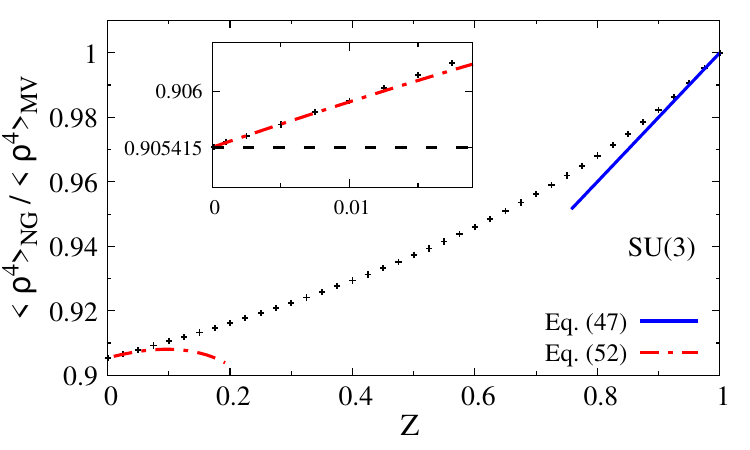}
	\end{center}
	\vspace*{-7mm}
	\caption[a]{(Left) Lattice spacing dependence of 
		the ratio of four-point function of color 
		charges in the non-Gaussian to the Gaussian theory 
		for different values of the renormalization factor 
		$Z$ for $L=1.77$ fm and $\mu=0.35$ GeV.
		(Right) Dependence of the same ratio 
		with $Z$. The inset plot shows that 
		$Z\rightarrow 0$ recovers the leading order 
		result (\eq{eq:nongauss_effect_SU3}) shown 
		as a dashed line.
		}
	\label{fig:SU3Ratiorho4FixedZ}
\end{figure}

The left panel of \fig{fig:SU3Ratiorho4FixedZ} shows the 
lattice spacing dependence of the ratio of the correlators of four-color charges
at the delta functions in the non-Gaussian and the Gaussian theories 
for different values of $Z$.
Smaller values of $Z$ lead to a larger deviation from the MV model. 
The curves at $Z=0.9999\sim 1$ and $Z=0.01$ 
are given by \eq{eq:scenario1pertres} and \eq{eq:scenario3nonpert}
respectively, which nicely reproduce the results of the non-perturbative calculations.
The right panel in \fig{fig:SU3Ratiorho4FixedZ}
shows the $Z$ dependence of the same ratio with points representing
the results from the non-perturbative calculation
together with the perturbative results from
\eq{eq:scenario1pertres} and \eq{eq:scenario3nonpert}.
We see that for both small and large non-Gaussian fluctuations, the
analytical results are in good agreement with the non-perturbative ones.
It is also shown that the ratio converges to
the result given by \eq{eq:nongauss_effect_SU3} in 
the $Z\to0$ limit.

\section{Conclusions}

In this work, we studied the non-perturbative effects of the first 
(even C-parity) non-Gaussian correction to the Gaussian theory of the CGC 
in SU(2) and SU(3) color symmetry groups. Deviations from the 
MV model were quantified via the renormalization factor, $Z$. 
The couplings in the non-Gaussian small-$x$ action need to 
be renormalized in order to reproduce the two-point function 
of color charges in the Gaussian theory. 
We considered three different 
renormalization schemes to determine the couplings of the non-Gaussian action. 
New analytical expressions were presented in the regime of large 
non-Gaussian fluctuations in each renormalization scheme and 
these were compared to numerical results
where the renormalization equation, \eq{eq:renormMuSU2}, was solved numerically. 
Our results pointed out that the first two renormalization schemes 
always lead to a theory dominated by non-Gaussian fluctuations 
independent of the system size. This means that even larger 
systems end up being strongly affected by non-Gaussian corrections. 
Such a scenario is unlikely to happen, as one expects 
the validity of the MV model for larger systems. 
The third renormalization scheme, where $Z$ is fixed,
on the other hand, allows one to control the deviations 
from the MV model. 
The strength of the non-Gaussian correction to the 
MV model in physical observables is still an open question and 
deserves further investigation. 
The next step is to determine the values of $Z$ by considering  
experimental data to see to what extent a system deviates from 
the MV model.

The calculations shown here represent the first practical step towards 
making non-Gaussian initial conditions to the JIMWLK evolution equations. 
In addition, we showed that the initial distribution of color charges moves
away from a Gaussian once deviations from the MV model are considered.
The quartic term should affect the multiplicity distribution, 
especially in small collision systems, where 
non-Gaussian corrections are usually expected. 
That would change the fluctuations of the energy 
(or gluon) density in the initial condition for hydrodynamic simulations.
In particular, fluctuations of the initial energy density are 
important to determine spatial eccentricities~\cite{Dumitru:2012yr},
which can be related to flow harmonics and angular correlations in hydrodynamic 
simulations~\cite{Ollitrault:1992bk,Poskanzer:1998yz,Alver:2010gr,Qin:2010pf,Teaney:2010vd,Gardim:2011xv,Qiu:2011iv,Bozek:2012gr}. 
Such changes also apply to early time fluctuations of axial charge density in the glasma phase, 
which are given in terms of the divergence of the Chern-Simons current~\cite{Lappi:2017skr,Guerrero-Rodriguez:2019ids}.

Furthermore, as shown in~\cite{Dumitru:2011ax}, the inclusion of a quartic term 
in the weight function generates a correction to the correlator of 
two Wilson lines, $\langle V(x)V^{\dagger}(y) \rangle$, 
where $V(x)$ denotes a Wilson line.
For this reason, such initial conditions may be used to study  whether 
there exist differences between the JIMWLK evolution with and without 
assuming the Gaussian approximation, where all higher $n$-point function 
of Wilson lines can be related to $\langle V(x)V^{\dagger}(y) \rangle$~\cite{Iancu:2011nj}.

The calculations in this paper can be extended to
study the non-Gaussian effects on the two-particle correlation function, $C_{2}(p,q)$
in the double inclusive gluon production~\cite{Lappi:2009xa}
and the dipole operator, 
$D(r) \propto \langle V(x)V^{\dagger}(y) \rangle$, 
complementing the results in the dilute regime 
from~\cite{Dumitru:2011ax}. 
In particular, it has been shown~\cite{Dumitru:2011zz} that
at the perturbative level the quartic term 
generates an additional contribution of the same order in $N_c$ to 
$C_{2}(p,q)$ on top of the contribution from the Gaussian part 
of the action. 
Moreover, the non-Gaussian correction becomes of the same order 
in the mass number and is enhanced by a factor of 
$N_c^2-1$ compared to the Gaussian contribution 
if one considers the saturation scale as a cutoff for 
integrals over transverse momentum figuring in this quantity.
A non-perturbative calculation is needed to access how the 
effects of additional contributions from a non-Gaussian 
statistics change the result from the MV model to all 
orders of $1/\kappa_4$ in this case.
Works in these directions are ongoing.

\begin{acknowledgments}
We are grateful to Adrian Dumitru for many discussions 
about non-Gaussian corrections to the MV model, 
helpful comments, and careful reading of the manuscript. 
A.V.G. acknowledges the Brazilian funding agency FAPESP
for financial support through grants 2017/14974-8 and 2018/23677-0.
Y. N. acknowledges the support by the
Grants-in-Aid for Scientific Research from JSPS (JP17K05448).

\end{acknowledgments}


\begin{thebibliography}{99}


\bibitem{CGC.review.new}
E.~Iancu and R.~Venugopalan,
In *Hwa, R.C. (ed.) et al.: Quark gluon plasma* 249-3363,
[hep-ph/0303204];
F.~Gelis, E.~Iancu, J.~Jalilian-Marian and R.~Venugopalan,
Ann.\ Rev.\ Nucl.\ Part.\ Sci.\  {\bf 60}, 463 (2010).


\bibitem{CGC.effective.theory}
Y.~V.~Kovchegov,
Phys.\ Rev.\ D {\bf 60}, 034008 (1999);
H.~Weigert,
Prog.\ Part.\ Nucl.\ Phys.\  {\bf 55}, 461 (2005);
J.~P.~Blaizot, F.~Gelis and R.~Venugopalan,
Nucl.\ Phys.\ A {\bf 743}, 13 (2004);
J.~P.~Blaizot, F.~Gelis and R.~Venugopalan,
Nucl.\ Phys.\ A {\bf 743}, 57 (2004);
H.~Weigert,
Prog.\ Part.\ Nucl.\ Phys.\  {\bf 55}, 461 (2005).


\bibitem{JalilianMarian:1996xn} 
J.~Jalilian-Marian, A.~Kovner, L.~D.~McLerran and H.~Weigert,
Phys.\ Rev.\ D {\bf 55}, 5414 (1997).

\bibitem{JalilianMarian:1997jx} 
J.~Jalilian-Marian, A.~Kovner, A.~Leonidov and H.~Weigert,
Nucl.\ Phys.\ B {\bf 504}, 415 (1997).

\bibitem{JalilianMarian:1997gr} 
J.~Jalilian-Marian, A.~Kovner, A.~Leonidov and H.~Weigert,
Phys.\ Rev.\ D {\bf 59}, 014014 (1998).

\bibitem{JalilianMarian:1997dw} 
J.~Jalilian-Marian, A.~Kovner and H.~Weigert,
Phys.\ Rev.\ D {\bf 59}, 014015 (1998).

\bibitem{JalilianMarian:1998cb} 
J.~Jalilian-Marian, A.~Kovner, A.~Leonidov and H.~Weigert,
Phys.\ Rev.\ D {\bf 59}, 034007 (1999);
Erratum: [Phys.\ Rev.\ D {\bf 59}, 099903 (1999)].

\bibitem{Kovner:1999bj} 
A.~Kovner and J.~G.~Milhano,
Phys.\ Rev.\ D {\bf 61}, 014012 (2000).

\bibitem{Kovner:2000pt} 
A.~Kovner, J.~G.~Milhano and H.~Weigert,
Phys.\ Rev.\ D {\bf 62}, 114005 (2000).

\bibitem{Iancu:2000hn} 
E.~Iancu, A.~Leonidov and L.~D.~McLerran,
Nucl.\ Phys.\ A {\bf 692}, 583 (2001).

\bibitem{Iancu:2001ad} 
E.~Iancu, A.~Leonidov and L.~D.~McLerran,
Phys.\ Lett.\ B {\bf 510}, 133 (2001).

\bibitem{Ferreiro:2001qy} 
E.~Ferreiro, E.~Iancu, A.~Leonidov and L.~McLerran,
Nucl.\ Phys.\ A {\bf 703}, 489 (2002).


\bibitem{CGC.Raju.McLerran}
L.~D.~McLerran and R.~Venugopalan,
Phys.\ Rev.\ D {\bf 49}, 2233 (1994),
Phys.\ Rev.\ D {\bf 49}, 3352 (1994),
Phys.\ Rev.\ D {\bf 50}, 2225 (1994).

\bibitem{Lam:2001ax} 
C.~S.~Lam and G.~Mahlon,
Phys.\ Rev.\ D {\bf 64}, 016004 (2001).


\bibitem{Dumitru:2011vk} 
A.~Dumitru, J.~Jalilian-Marian, T.~Lappi, B.~Schenke and R.~Venugopalan,
Phys.\ Lett.\ B {\bf 706}, 219 (2011).


\bibitem{Marquet:2007vb} 
C.~Marquet,
Nucl.\ Phys.\ A {\bf 796}, 41 (2007).


\bibitem{Dominguez:2011wm} 
F.~Dominguez, C.~Marquet, B.~W.~Xiao and F.~Yuan,
Phys.\ Rev.\ D {\bf 83}, 105005 (2011).


\bibitem{Lappi:2015vta}
T.~Lappi, B.~Schenke, S.~Schlichting and R.~Venugopalan,
JHEP {\bf 1601}, 061 (2016).




\bibitem{Jeon:2004rk}
S.~Jeon and R.~Venugopalan,
Phys.\ Rev.\ D {\bf 70}, 105012 (2004);
Phys.\ Rev.\ D {\bf 71}, 125003 (2005).

\bibitem{Dumitru:2011zz} 
A.~Dumitru, J.~Jalilian-Marian and E.~Petreska,
Phys.\ Rev.\ D {\bf 84}, 014018 (2011).

\bibitem{Dumitru:2011ax} 
A.~Dumitru and E.~Petreska,
Nucl.\ Phys.\ A {\bf 879}, 59 (2012).

\bibitem{Dumitru:2012tw} 
A.~Dumitru and E.~Petreska,
arXiv:1209.4105 [hep-ph].



\bibitem{Kovner:2010xk} 
A.~Kovner and M.~Lublinsky,
Phys.\ Rev.\ D {\bf 83}, 034017 (2011).



\bibitem{GolecBiernat:1998js} 
K.~J.~Golec-Biernat and M.~Wusthoff,
Phys.\ Rev.\ D {\bf 59}, 014017 (1998).


\bibitem{Dumitru:2010mv}
A.~Dumitru and J.~Jalilian-Marian,
Phys. Rev. D \textbf{81}, 094015 (2010).


\bibitem{Dumitru_priv_comm}{
A.~Dumitru, private communication.
}

\bibitem{Krasnitz:1998ns} 
A.~Krasnitz and R.~Venugopalan,
Nucl.\ Phys.\ B {\bf 557}, 237 (1999).


\bibitem{Dumitru:2012yr}
  A.~Dumitru and Y.~Nara,
  Phys.\ Rev.\ C {\bf 85}, 034907 (2012).


\bibitem{Ollitrault:1992bk} 
J.~Y.~Ollitrault,
Phys.\ Rev.\ D {\bf 46}, 229 (1992).

\bibitem{Poskanzer:1998yz} 
A.~M.~Poskanzer and S.~A.~Voloshin,
Phys.\ Rev.\ C {\bf 58}, 1671 (1998).

\bibitem{Alver:2010gr} 
B.~Alver and G.~Roland,
Phys.\ Rev.\ C {\bf 81}, 054905 (2010)
Erratum: [Phys.\ Rev.\ C {\bf 82}, 039903 (2010)].

\bibitem{Qin:2010pf} 
G.~Y.~Qin, H.~Petersen, S.~A.~Bass and B.~Muller,
Phys.\ Rev.\ C {\bf 82}, 064903 (2010).

\bibitem{Teaney:2010vd} 
D.~Teaney and L.~Yan,
Phys.\ Rev.\ C {\bf 83}, 064904 (2011).

\bibitem{Gardim:2011xv} 
F.~G.~Gardim, F.~Grassi, M.~Luzum and J.~Y.~Ollitrault,
Phys.\ Rev.\ C {\bf 85}, 024908 (2012).

\bibitem{Qiu:2011iv} 
Z.~Qiu and U.~W.~Heinz,
Phys.\ Rev.\ C {\bf 84}, 024911 (2011).

\bibitem{Bozek:2012gr} 
P.~Bozek and W.~Broniowski,
Phys.\ Lett.\ B {\bf 718}, 1557 (2013).


\bibitem{Lappi:2017skr} 
T.~Lappi and S.~Schlichting,
Phys.\ Rev.\ D {\bf 97}, no. 3, 034034 (2018).

\bibitem{Guerrero-Rodriguez:2019ids} 
P.~Guerrero-Rodr\'{i}guez,
JHEP {\bf 1908}, 026 (2019).



\bibitem{Iancu:2011nj}
E.~Iancu and D.~N.~Triantafyllopoulos,
JHEP {\bf 1204}, 025 (2012).


\bibitem{Lappi:2009xa} 
T.~Lappi, S.~Srednyak and R.~Venugopalan,
JHEP {\bf 1001}, 066 (2010).



\end{thebibliography}
\end{document}